\documentclass[preprint]{aastex} 

\slugcomment{Revised \& accepted: 20/08/2003}
\shorttitle{2MASS Ultracool Dwarfs}
\shortauthors{Reid et al.}

\begin{document}
\def\pedant{Mart{\'{\i}}n }
\def\etal{{\sl et al.}}
\def\etall{{\sl et al. }}
\def\pma{$\arcsec$~yr$^{-1}$ }
\def\kms{km~s$^{-1}$ }
\def\msun{$M_{\odot}$}
\def\rsun{$R_{\odot}$}
\def\lsun{$L_{\odot}$}
\def\halpha{H$\alpha$}
\def\hbeta{H$\beta$}
\def\hgama{H$\gamma$}
\def\hdelta{H$\delta$}
\def\Teff{T$_{eff}$}
\def\logg{$log_g$}

\title{Meeting the Cool Neighbors VII: Spectroscopy of faint, red NLTT dwarfs }

\author{I. Neill Reid\altaffilmark{1}}
\affil{Space Telescope Science Institute, 3700 San Martin Drive, Baltimore, MD 
21218; and
Department of Physics and Astronomy, University of Pennsylvania, 209 South 33rd 
Street,
Philadelphia, PA  19104; inr@stsci.edu}

\author{Kelle L. Cruz\altaffilmark{1}, Peter Allen\altaffilmark{1}, F. Mungall\altaffilmark{1}}
\affil{Department of Physics and Astronomy, University of Pennsylvania, 
209 South 33rd Street, Philadelphia, PA  19104; kelle@sas.upenn.edu}

\author{D. Kilkenny}
\affil{South African Astronomical Observatory, P.O. Box 9, Observatory 7935, 
South Africa}

\author{James Liebert\altaffilmark{1}}
\affil{Department of Astronomy and Steward Observatory, 
University of Arizona, Tucson, AZ 85721; liebert@as.arizona.edu}

\author{Suzanne L. Hawley, Oliver J. Fraser, Kevin R. Covey}
\affil{Department of Astronomy, University of Washington, Box 351580, Seattle, WA 98195}

\author{Patrick Lowrance\altaffilmark{1}}
\affil{IPAC, MS 100-22 Caltech, 770 S. Wilson Ave., Pasadena, CA 91125}

\altaffiltext{1}{Visiting Astronomer, Kitt Peak National Observatory, NOAO, 
which is operated
by AURA under cooperative agreement with the NSF.}

\begin{abstract}
We present low-resolution optical spectroscopy and BVRI photometry of 453 candidate 
nearby stars drawn from the NLTT proper motion catalogue. The stars were
selected based on optical/near-infrared colours, derived by combining the NLTT 
photographic data with photometry from the 2MASS Second Incremental Data Release. Based on
the derived photometric and spectroscopic parallaxes, we identify 111 stars 
as lying within
20 parsecs of the Sun, including 9 stars with formal distance estimates of less
than 10 parsecs. A further 53 stars have distance estimates within 1$\sigma$ of
our 20-parsec limit.
Almost all of those stars are additions to the nearby star
census. In total, our NLTT-based survey has so far identified 496 stars likely
to be within 20 parsecs, of which 195 are additions to nearby-star catalogues.
Most of the newly-identified nearby stars have spectral types between M4
and M8. 

\end{abstract}


\mbox{}

\section{Introduction}

Recent years have seen renewed interest in the stellar populations in the 
immediate Solar Neighbourhood, stimulated in part by the discovery of 
significant numbers of extrasolar planetary systems. Understanding the frequency 
and distribution of those systems as a function of stellar properties is clearly 
fundamental in understanding the formation mechanism(s), and the nearest stars 
provide the best observational opportunities for studying such issues. A vital 
step in this process is compiling a reliable census of the local stars and brown 
dwarfs. The Hipparcos survey (ESA, 1997) provides accurate statistics for 
earlier type stars (spectral types A to K), albeit of limited completeness 
beyond d$\sim$30 parsecs. However, catalogues of low-luminosity later-type 
dwarfs become woefully incomplete at distances of only 10 to 15 parsecs from the 
Sun (Reid, Gizis \& Hawley, 2002 - PMSU4; Henry \etal, 2002), yet those systems 
make a
substantial numerical contribution to the local stellar population.

As part of the NASA/NSF NStars initiative, 
we are currently undertaking a large-scale survey which aims to improve 
significantly the Solar Neighbourhood census of late-type dwarfs. With effective 
temperatures T$_{eff} < 4000$K, these stars and brown dwarfs emit the bulk of their radiative 
flux at near-infrared wavelengths. We are therefore using JHK$_S$ photometry 
provided by the 2-Micron All Sky Survey (2MASS - Skrutskie \etal, 1997), both in 
isolation and in combination with other surveys, to search for previously 
unrecognised M and L dwarfs within 20 parsecs of the Sun. 

Ultracool dwarfs (spectral types $\ge$ M7) have distinctive (J-K$_S$) colours, 
and nearby candidates can be identified directly from the 2MASS 
photometry (as described by Cruz \etal, 2003, Paper V). Earlier type M dwarfs 
are less distinctive at near-infrared wavelengths, requiring that we supplement 
the 2MASS data. Proper motions have long proven an effective means of 
identifying nearby stars, particularly Luyten's Palomar Schmidt surveys. The 
most extensive such catalogue is the NLTT (Luyten, 1980), including over 58,000 
stars with $\mu \ge 0.18$\pma (corresponding to a tangential velocity V$_{tan} > 
19$\kms at 20 parsecs). The NLTT does not extend to extremely faint magnitudes. 
Salim \& Gould (2002) have undertaken a recent thorough analysis, and conclude 
that the catalogue becomes significantly incomplete at m$_r > 17$. However, the 
catalogue is more than adequate for our purposes, since this limit corresponds 
to the apparent magnitude of an M7 dwarf at 20 parsecs. We therefore expect 
analysis of the NLTT catalogue to provide additional mid- to late-type M dwarfs 
(M3-M7) within 20 parsecs, complementing the ultracool dwarfs identified directly from 
the 2MASS photometry. 

Our analysis is restricted to the $\sim48\%$ of the sky covered by the 2MASS 
Second Incremental Release (Skrutskie \etal, 2000; hereinafter, 2M2nd). 
Paper I in this series (Reid \& Cruz, 2002) 
described how we cross-referenced the NLTT catalogue against that database and 
identified over 23,000 proper motion stars with 2MASS counterparts within 10 
arcseconds of the predicted position (J2000, epoch 1998.0). Using the
(m$_r$-K$_S$) colours for those stars, where m$_r$ is from the NLTT, 
we selected a sample of 1245 stars likely to be within 20 
parsecs of the Sun. This sample is defined as NLTT Sample 1 
(NLTT1)\footnote{ A subsequent paper 
in this series will discuss follow-up observations of two additional NLTT 
samples: NLTT2, compiled by searching for potential 
2MASS counterparts within 60 
arcseconds of predicted NLTT positions; and NLTT3, drawn from the rNLTT 
Salim \& Gould (2002).}. As discussed further below, eight of those candidates 
have since been eliminated as spurious, reducing the total to 1237 stars.
Four hundred and sixty-nine of the NLTT1 stars are known nearby dwarfs, with 
published photometric or spectroscopic data (as summarised in Paper I). Paper II 
(Reid, Kilkenny \& Cruz, 2002) presented BVRI photometry for a further 180 
stars,
while Paper III (Cruz \etal, 2002) discussed optical spectroscopy of 
over 50 of the fainter candidates. We have continued to obtain follow-up 
observations of this sample, and this paper presents optical spectroscopy and 
photometry of 453 M dwarfs. With the addition of those data, we have 
distance estimates for 1126 stars from NLTT Sample 1, including all 
378 stars with m$_r > 14.5$. 

The paper is organised as follows: \S2 reviews the target selection and
describes the spectroscopic and photometric observations; \S3 discusses the 
bandstrengths measured for the more prominent spectral features, and presents 
the derived spectral types and distance estimates; \S4 discusses a number of the 
more interesting stars; and \S5 summarises the overall results and our 
conclusions.

\section{Spectroscopic and photometric observations}

\subsection {Target priorities: late-type dwarfs in the NLTT}

Proper motion surveys are kinematically defined samples: the effective sampling 
volume is determined by the average tangential velocity, $\langle V_T \rangle$, 
of the particular stellar population surveyed. In the case of the Galactic disk, 
$\langle V_T \rangle \approx 37$ \kms (Reid, 1997), so the characteristic distance 
limit of stars in the NLTT catalogue ($\mu > 0.18$\pma) is $\sim43$ parsecs. All 
NLTT stars are drawn from the same volume, so stars at fainter apparent 
magnitudes generally also have fainter absolute magnitudes. 
The local stellar census is least complete for late-type dwarfs, and those low 
luminosity systems are the highest priority targets of our NStars survey. We 
have therefore concentrated our follow-up spectroscopic observations on the 
faintest stars in NLTT1. 

In particular, we have targeted stars in NLTT1 with $m_r > 14.5$, 
corresponding to $M_r > 13$ or spectral types 
later than $\approx$M3 for distances $d<20$ parsecs. As noted above, 
the initial sample defined in paper I included 1245 stars. More
detailed scrutiny has shown that eight of those stars were paired with
2MASS data for other stars within the 10-arcsecond search radius. 
None of the eight NLTT dwarfs meet our (m$_r$, (m$_r$-K$_S$))
criteria when matched correctly (two are white dwarfs), reducing the total sample to 1237 stars,
of which 378 have m$_r > 14.5$.  
Forty-four of the faint subset have published photometry and 
are included in Paper I, while a further 32 are discussed in Paper III
(supplementary observations of fifteen stars from Paper III are
presented here). Thirteen stars are either well-known late-type dwarfs or 
are sufficiently red at $(J-K_S)$ that they are included in the 2MASS-selected 
sample discussed in Paper V. Relevant properties of those stars
are summarised in Table 1. Spectroscopic observations of the remaining 288
NLTT1 stars with m$_r > 14.5$ are presented in this paper together with
data for a number of brighter stars.

Broadband photometric colours provide an alternative means of estimating 
distances and luminosities of late-type dwarfs and, as discussed in Paper II, we 
are using this technique for follow-up observations of many of 
the brighter stars (m$_r < 14.5$) in the 
NLTT 1 sample. We have continued our program of follow-up observations, and BVRI 
data for a further 96 stars are presented in this paper.

\subsection {Spectroscopy}

We have obtained low-resolution optical spectroscopy of 357 NLTT dwarfs, 
covering the 6000 to 10000 \AA\ wavelength range. Astrometry and 2MASS/NLTT 
photometry for all targets are given in Table 2. The observations were obtained 
by a variety of observers at Kitt Peak National Observatory, Cerro Tololo 
Interamerican Observatory and Apache Point Observatory on the following 
occasions.

Cruz, Reid and Allen used the GoldCam spectrograph on the Kitt Peak 2.1m on 2001 
November 1{--}5, 2002 July 2{--}7 and 2003 March 12{--}16. The weather was 
generally photometric 
with moderate seeing (1\farcs5). On each occasion, we employed the 400 
line mm$^{-1}$ grating, blazed at 8000\AA, with an OG550 order-blocking filter.
Observations were made using a 2\farcs0 slit. The spectra have a resolution 
of 5.5\AA. The Ford 3K$\times$1K CCD used in this spectrograph produces significant fringing 
redward of $\sim8200\AA$. None of the more important spectral features discussed 
in this paper lie at those wavelengths, however. 

Observations were also made by Cruz, Liebert, C. Cooper  and 
N. Gorlova at Kitt Peak on 2001 July 
13-23, using the RC spectrograph on the 4m Mayall telescope. An OG530 filter was 
employed as an order-blocking filter, and the 317 line mm$^{-1}$ grating coupled 
with a 1\farcs5 slitwidth gave a spectral resolution of 5.6\AA\ (2 pixels). The 
weather was generally favourable throughout the run, with partly cloudy 
conditions and a wide range of seeing conditions. Further observations using a 
similar instrumental setup (an OG550 order-blocking filter replaced the OC530) 
were made by Liebert and Lowrance on 2002 January 21-24. Seeing was  
better, with a 1\farcs0 slit used on January 21, 22 and 24 (spectral resolution 
4.7\AA), although the 1\farcs5 slit was employed on January 23. Additional
observations were obtained by Reid, Cruz, Liebert and Mungall using the Mayall
telescope with the Multi-Aperture Red Spectrometer (MARS) on 2002 September 24-28
and 2003 July 8-13. The VG8050-450 grism was employed with a 2\arcsec\ slit, giving an
resolution of 7\AA (3.5 pixels). High cirrus was present during most of the
July 2003 observations.

A number of southern targets were observed on 2002 January 26{--}30 by Reid and 
Cruz using the RC spectrograph on the CTIO 1.5-metre. Cruz and Mungall obtained
further observations with this telescope and spectrograph on 2003 May 14{--}17.
Seeing was between 0\farcs7 and 2\farcs0 throughout both runs. 
We combined the 500 line mm$^{-1}$ grating, 
blazed at 8000\AA, with a 1\farcs5 slit-width to give spectra with a resolution 
of 6.5\AA\ (3 pixels). As with the 2.1-metre observations, the Loral 1K CCD is 
subject to significant fringing beyond $\sim8000$\AA. 

Finally, thirty stars were observed by Hawley, Fraser and Covey using 
the Double Imaging Sectrograph (DIS II) on 
the 3.5-metre telescope at Apache Point Observatory.  
Spectra were obtained on 10 April, 14 May and 30 May 2002. The 
medium resolution (300 line  mm$^{-1}$) grating 
 was employed on the red camera, giving a dispersion of 3.15 \AA pix$^{-1}$. 
Seeing was typically 1\farcs5, 
and a 1\farcs5 slitwidth was employed for these observations, giving a 
resolution of 7.3\AA\ (2 pixels). 

All spectra were bias-subtracted, flat-fielded, corrected for bad pixels, 
extracted and wavelength- and flux-calibrated using the standard IRAF packages 
CCDPROC and DOSLIT. The wavelength calibration was determined from He-Ne-AR arcs 
taken at the start of each night. The flux calibration was determined using the 
standard stars HD 19445, HD 84937, Ross 640, G191-B2b, L1363-3, HZ 4 and Feige 
34 (Oke \& Gunn, 1983; Hamuy \etal, 1994). At least one flux standard  
was observed each night, and a comparison of repeated independently-calibrated 
observations of individual program stars indicates that the derived spectral 
energy distributions are consistent to better than 5\% over the full wavelength 
range. This is more than adequate for the purposes of this program.

M dwarf spectra are dominated by molecular bands. We have measured bandstrengths 
for a number of those features using the techniques outlined in Paper III. The 
derived indices for TiO (7020\AA\ band), CaH (6300 and 6800\AA), CaOH (6250 
\AA), VO (7300 \AA) and H$\alpha$ are listed in Table 3. Our measurements of 
the standard stars show excellent agreement with previously published data, with
no systematic offset and typical dispersions of $\pm0.01$. Moreover, ten stars
from the present sample have independent observations listed in Table 5 of 
Paper III. Direct comparison with those measurements gives mean differences
$| \Delta | < 0.01$ and rms dispersions of $\sigma < 0.02$. 

\subsection {Photometry}

The photometric follow-up observations were obtained by D. Kilkenny between 2001 
December and 2002 July, using the St. Andrews photometer on the 1 metre 
telescope at the Sutherland station of the South African Astronomical 
Observatory. The photometer is equipped with a Hamamatsu R943-02 GaAs 
photomultiplier, and observations were made using a Johnson-Cousins filter set. 
Most observations were made through a 21\arcsec\ diameter aperture, with a 
31\arcsec\ aperture employed for conditions of poor seeing. The relatively large 
apertures lead to our photometry including contributions from other stars in a 
few cases, as noted below.

The observations were reduced using identical methods to those outlined in Paper 
II (see also Kilkenny \etal, 1998). In general, we obtained B-band data for only 
the brightest stars. The program star photometry was calibrated using 
observations of both E-region standards (Cousins, 1973; Menzies \etal, 1989) and 
redder standards from Kilkenny \etal\ Figure 1 shows the residuals in V, (B-V), 
(V-R) and (V-I) derived from our observations of the latter stars. In each case, 
the overall rms uncertainties are better than 1\%. None of the NLTT stars 
discussed here have previous observations. However, Paper II presented an 
extensive comparison between SAAO and literature data for late-type dwarfs, 
demonstrating that our photometry is accurate and consistent with standard 
photometric systems to better than 1\%. We note that the uncertainties in
the derived photometric parallaxes are dominated by the intrinsic width of the
main sequence, which leads to dispersions $\sigma = 0.25$ to 0.4 magnitudes
in the mean calibrating relations (Paper I). 

Table 4 presents the results of our photometry of the NLTT stars. In addition to 
the optical data, we have listed the positions and near-infrared photometry from 
the 2M2nd\footnote { Note that slight changes in both 
the astrometry and photometry of these objects may be present in the final 2MASS 
data release.} Table 4 gives the number of observations of each star and, in 
cases with multiple observations, 
the rms dispersion in the derived mean magnitudes and colours. Based on these 
results and past 
experience, we expect that the typical uncertainties in our VRI photometry are 
$<2\%$, even for stars with single 
observations. A few stars show larger night-to-night residuals, primarily in 
magnitude 
rather than colour. We interpret these higher residuals as indications of 
variability at the $<0.1$ 
magnitude level. The relevant stars are identified in Table 4.

\section {Spectral types, distances and absolute magnitudes}

Our primary goal is identifying late-type dwarfs which are likely to lie within 
20 parsecs of the Sun. The spectroscopic and photometric data that we have 
obtained can provide distance estimates. However, it is important to bear in 
mind that the main sequence has a significant intrinsic width, while our distance 
estimates are calibrated against the mean properties of the local disk 
population. Trigonometric parallax measurements remain the most reliable method 
of distance determination for individual stars, and eventually all of the 
nearby-star candidates identified in our project should be targetted by parallax 
programs.

\subsection {Spectroscopic parallaxes}

We expect that the majority of the stars in the NLTT1 sample are drawn from the 
disk population. The narrowband indices measured from our low-resolution spectra 
provide a means of checking that assumption. Cool metal-poor subdwarfs show 
enhanced CaH absorption relative to TiO absorption (Jones, 1973; Mould, 1976). Gizis (1997) 
has calibrated this behaviour using the same CaH indices which we have measured 
in this paper. Figure 2 compares the CaH2/TiO5 and CaH3/TiO5 distributions for 
the NLTT stars against reference data for disk dwarfs (from Reid \etal, 1995 - 
PMSU1) and intermediate and extreme subdwarfs (from Gizis, 1997). As discussed further below,
all save one of the NLTT dwarfs have CaH/TiO strengths consistent with their being members of the disk 
population. 

We have used the narrowband indices listed in Table 3 to derive spectral types 
and absolute magnitude estimates for the 357 stars with spectroscopic data. The 
spectral types are calibrated primarily using the TiO5 index, following the 
relation derived in Paper III. 
The 7020\AA\ TiO bandhead saturates at spectral type M6, leading to the breakdown 
of this calibration method, with TiO5 decreasing in strength at later types. 
Fortunately, the VO-a index provides an alternative 
means of calibrating spectral types earlier than M8 (there are no later type 
stars in the 
present sample). In general, the TiO5 and VO-a spectral type calibrations agree 
to better than a
spectral class, so discordant classifications allow us to identify 
the few M6-M8 dwarfs in the present sample. The spectral classification of 
those stars rests on visual examination, as discussed in Paper V. Indeed,
we have inspected all of the spectra and verified the classifications listed in Table 3.
Thirteen of the fifteen stars from Paper III are included in Table 3, and the
spectral types differ by less than 0.5 classes; in most cases, the two
measurements are in exact agreement.

Narrowband indices also provide a means of estimating absolute magnitudes. We 
have employed 
the CaH2, CaOH and TiO5 indices for this purpose, utilising the calibrating 
relations defined 
in Paper III. As outlined in that paper, all three calibrations show the same 
morphology, with 
a clear discontinuity between upper and lower branches at $8 < M_J < 9$ 
(spectral type 
$\approx$M3). This feature appears at the same effective location in the main 
sequence in 
all colours (Paper I, Williams \etal, 2002). In Paper III we fitted separately 
the upper 
and lower branches of the colour-magnitude diagram (CMD) for each narrowband 
index. Since 
M$_J$ is inherently ambiguous for stars falling in that region of the CMD, we 
have simply 
averaged the absolute magnitudes derived from the upper and lower calibrations 
for a
 particular index, and assign an uncertainty of $\pm0.4$ magnitudes. 

Absolute  magnitudes and 
uncertainties for stars with spectral types earlier than M6 are derived by 
averaging the three individual estimates (TiO5, CaH2 and CaOH).  
The uncertainties of $\pm0.02$ in the measured bandstrengths correspond to typical 
uncertainties of $\pm 0.14$ in M$_J$, generally less than the width of the 
main sequence over the range covered by this calibration. 
Both CaH2 and CaOH follow TiO5 in showing a reversal in behaviour at spectral 
type $\approx$M6. 
As a result, we cannot use these indices to estimate absolute magnitudes for 
later type dwarfs. 
Instead, we have estimated absolute magnitudes for those stars using the 
spectral-type/M$_J$ calibration defined in Paper V. 
The derived absolute magnitudes are listed in Table 2.

The one possible non-disk star is LP 381-49, which lies close to the sdM sequence in the
(TiO5, CaH2) plane (at [0.32, 0.25]) and on the lower edge of the disk distribution in the
(TiO5, CaH2) plane (at [0.32, 0.56]). Our spectrum of that star, obtained with the 
CTIO 1.5-metre, is compared to the spectral standards Gl 83.1 (M4.5) and Gl 65 (M5.5) 
are shown in Figure 3. The signal-to-noise in our spectrum is only moderate, but
H$\alpha$ is clearly present in emission. 
If confirmed as an intermediate subdwarf, then the star is likely to be even closer than
our current estimate of 14 parsecs.

\subsection {Photometric parallaxes}

We have followed the approach outlined in Papers I and II to estimate 
photometric parallaxes 
for the NLTT dwarfs with broadband photometry. As in those previous papers, we 
combine estimates 
based on three colour indices - (V-I), (V-K) and (I-J) - using the relations 
defined in Paper I. 
Broadband CMDs also show the discontinuity at $\approx$M3 evident in the 
narrowband CMDs, and the 
associated uncertainties in absolute magnitude increase accordingly. Elsewhere, 
typical 
uncertainties for an individual photometric parallax are between $\pm0.25$ and 
$\pm0.35$ 
magnitudes. Recognising the intrinsic dispersion in the main sequence, 
we set a 
lower limit of $\pm0.25$ mag on the uncertainty associated with the 
combined, mean absolute magnitude.

Our BVRI data are derived from aperture photometry, and, as noted is $\S2.2$, in 
a few cases 
the aperture includes additional stars, either background field stars or genuine 
binary 
companions. The five stars affected by this problem are identified in Table 4, 
and we used 
images from 2MASS and the DSS to estimate the appropriate adjustments to the V 
magnitude 
(we assume that the optical colours are largely unaffected). In one case, G 274-
24 (or GJ 
2022AC), the offending star, an equal-magnitude binary companion, is 
sufficiently close that 
2MASS also fails to resolve the system. Jao \etal (2002) have used higher 
spatial-resolution imaging to determine the relative magnitudes of the 
components in this system, and we have adjusted both the optical and near-
infrared data accordingly. 
The photometric parallaxes for these five systems are correspondingly less
certain than those of the other NLTT stars.

Our final 
estimates of the absolute magnitudes and distances of the 96 NLTT dwarfs with 
BVRI observations are given in Table 4. Two stars have spectroscopic observations
from Paper III: LP 768-113 and G 75-35. In the former case, our spectroscopic
distance estimate is 17.5 parsecs, in exact agreement with the photometric 
estimate. G75-35, with spectral type M4, lies near the break in the main-sequence
and we gave two distance estimates in Paper III - $11.2\pm0.8$ parsecs and 
$17.3\pm1.8$ parsecs. The former is consistent with our current
 photometric estimate of $12.4\pm1.8$ parsecs. As with all stars with spectral types M3-M4, 
trigonometric parallax data are required for definitive distance measurements.

\section {Discussion}

\subsection {Distances}

The distance distributions of both spectroscopic and photometric samples are 
shown in Figure 4. 
Both show the same broad overall trend, with earlier-type/bluer stars 
lying at larger 
distances. This is as expected, since, in both cases, the targets are drawn 
primarily from a limited range 
of apparent magnitude - $14.5 < m_r < 18$ for the spectroscopic sample, and $11 
< m_r < 14.5$ 
for the photometric sample. Lower luminosity stars in
each sample are, of necessity, at smaller distances. 

Of the stars included in this paper, over 100 have formal distance estimates 
of less than 
20 parsecs. 36 of the 96 stars with BVRI photometry have mean photometric 
parallaxes 
$\pi_{ph} \ge 0\farcs05$; a further 16 have distance estimates within 1$\sigma$ 
of our 20 
parsec limit. Similarly, 75 of the 357 stars with spectroscopic parallaxes have 
inferred distances of less than 20 parsecs, 
while a further 37 lie within 1$\sigma$ of the limit. Few, if any, of these
stars have appeared in previous catalogues of the local stellar population, 
although we note that several are
currently under study by the CTIOPI program (Henry \etal, 2002).

\subsection {Stars of particular interest}

\begin{description}
\item[The nearest stars:] Four stars in the spectroscopic sample and five stars 
in the
photometric sample have formal distance estimates of less than 10 parsecs. 
From the spectroscopic sample, these are: \\
LP 71-81 (d=7.1 pc, M4.5), LP 206-11 (7.8 pc, M6.5), LP 349-25 (8.5 pc, M7.5) 
and
LP 30-55 (d=9.8 pc, M4.5); \\
from the photometric sample, 
LP 876-10 (7.2 pc, (V-I)=2.94, $\sim$M4.5), LP 984-92
(8.3 pc, (V-I)=3.00, $\sim$M4.5), LP 991-84 (8.3 pc, (V-I)=2.94, $\sim$M5), 
LP 869-26 (9.2 pc, (V-I)=3.19, $\sim$M4.5/M5) and LP 869-19 (9.8 pc, (V-I)=2.91,
$\sim$M4.5). \\
Several of these stars lie close to the break in the disk main-sequence, where
photometric and spectroscopic parallaxes are particularly uncertain.
Nonetheless, all nine are high priority candidates for trigonometric parallax programs.

\item[Strong H$\alpha$ stars:] Many of the NLTT M dwarfs have detectable
emission at H$\alpha$, and the statistics for the full sample will be discussed 
by Liebert \etal (in prep.). However, several stars stand out as having 
particularly 
strong emission. These include LP 209-2 (EW=10.1\AA), LP 222-65 (13.8\AA), LP 
349-25 (13.5\AA),
LP 373-35 (10.7\AA), LP 423-31 (26.0\AA; see also paper V); LP 763-3 (11.9\AA), 
LP 763-61 (11.4\AA), LP 800-58 (12.9\AA), LP 820-16 (12.2\AA) and LP 833-4 
(17.9\AA)
In addition, we re-observed the M9 dwarf LP 647-13 (d$\sim$10.5 pc) in the 
course
of our July 2001 KPNO observing run. Our original observations, from September 
2000, shows
prominent H$\alpha$ emission, equivalent width 10.2\AA; the July 2001 
observation has significantly
stronger emission, EW=36.2\AA, suggesting that we happened to catch it during a 
flare.
There is no evidence in our spectrum for emission due to helium or alkaline elements.

\end{description}

\subsection {The nearby star census}

With the addition of the observations described in this paper, we have 
quantitative distance estimates for 90\% of the stars in the NLTT1 sample. 
Figure 5 shows the (m$_r$, (m$_r$-K$_S$) and (J-H)/(H-K$_S$) distributions for 
the 111 nearby-star candidates from that sample which still lack such data. All 
of those stars are brighter than m$_r$=14.5 and have colours consistent with 
early- or mid-type M dwarfs. In addition to those stars, our
nearby census must also 
include contributions from stars in the NLTT2 and NLTT3 samples. Nonetheless, we 
can undertake a preliminary assessment of the likely impact of our study on the 
statistics of local low-mass stars.

The current standard reference catalogue of nearby M dwarf is the third 
Catalogue of
Nearby Stars (Gliese \& Jahreiss, 1991: CNS3), supplemented by the extensive 
follow-up spectroscopy of the PMSU survey (PMSU1; Hawley, Gizis
\& Reid , 1996).
This catalogue lists $\sim2150$ M dwarfs, including 150 unresolved 
companions, most of which lie within 25 parsecs of the
Sun. The distance limit for completeness, however, decreases sharply with
increasing spectral type, with d$_{lim} = 14$ pc for M3/M4 dwarfs,  d$_{lim} = 
10$ pc
for M5/M6 dwarfs and  d$_{lim} = 5$ pc for ultracool dwarfs 
(PMSU4). We have cross-referenced the PMSU dataset 
against both the 2M2nd and the NLTT catalogue.
 Nine hundred and forty-eight of the 1966 resolved M dwarfs in the PMSU fall 
within the area covered by the former database, including 530 stars, from 476 
systems, with
PMSU distance estimates of less than 20 parsecs. We can match the statistics for
that sample against the preliminary results from our NLTT-based survey. 

First, the PMSU sample includes 96 stars which do not appear in the NLTT, 
although
only 28 have formal distance estimates d$_{PMSU} < 20$ pc. Twenty-one of the 
latter 28 have 
spectral types earlier than M2. Most of the non-NLTT  stars are drawn from 
objective 
prism surveys (e.g. Stephenson, 1986; Sanduleak, 1976; Robertson, 1984), and 
have 
proper motions below the NLTT limit. A few stars, however, have $\mu 
> 0.18$\pma, but were not included in Luyten's LHS catalogue.
For example, amongst the d$_{PMSU} < 20$ pc stars, Vyssotsky 759 (1607+53)
has $\mu = 0.23$\pma, but is not listed in the NLTT.

We can use the PMSU stars included in the NLTT as a check on our (m$_r$, 
(m$_r$-K$_S$)) criteria, 
which are designed to select, with a comfortable margin, stars within
20 parecs of the Sun. Four hundred and sixty-eight of the 502 $d_{PMSU}<20$ pc. 
stars meet those criteria, as do 226 stars with $d_{PMSU} > 20$ pc (Figure 6).
Thus, almost 90\% of the 20-parsec PMSU stars are 
included in our sample\footnote{ Note that some of these stars are in the NLTT2 
and NLTT3 samples.}. We will discuss the 34 `missing' $d_{PMSU}<20$ pc stars in
more detail when we consider a complete analysis of the local luminosity 
function. For
the moment, we note that most of the stars have spectral types of M3 or M4, and
therefore lie near the break in the disk colour-magnitude relation (see Paper I).

Combining the results from Papers  I-III with the current paper, we have 
identified 496 NLTT M dwarfs as likely to be within 20 parsecs of the Sun.
One hundred and ninety-five
of those stars are not included in the CNS3 and are
additions to the local census, with most having spectral types 
later than M4. Figure 7 compares the J-band absolute magnitude distribution for 
the NLTT1 d$<$20 parsec stars against the PMSU-2M2nd 20-parsec 
sample (511 stars). Even though the NLTT sample is still
far from complete, it is clear that already we have more than doubled the number 
of known nearby dwarfs with M$_J > 9$ (M$_V > 13$, or spectral types later than M4).  
This is not surprising, given the limitations of the CNS3/PMSU dataset outlined 
above. 
These preliminary results indicate that our survey of NLTT stars offers the prospect 
of greatly improving the completeness of the local census of late-type dwarfs.

\section {Summary}

We have presented far-red spectroscopic observations and broadband BVRI 
photometry for 453 proper motion stars drawn from the NLTT catalogue. The stars 
were selected for observation based on their having a 
location in the ($m_r$, ($m_r-K_S$)) 
consistent with distances of less than 20 parsecs from the Sun. We 
have used narrowband spectroscopic indices, spectral types and broadband colours 
to derive spectroscopic and photometric parallaxes. Those more accurate distance 
estimates confirm that 111 lie within our sampling volume, while a further 53
have distances within 1$\sigma$ of the formal distance limit. 
We expect that the completion of our NLTT survey will lead to a substantial
increase in the number of known nearby late-type dwarfs
and a corresponding increase in our knowledge of the properties of
those objects. Combined with the ultracool dwarfs from the 2MASS
survey described in Paper V, we anticipate obtaining a much improved, 
statistically-robust measurement of the luminosity function of 
Solar Neighbourhood late-M and L dwarfs.

\acknowledgments
This research was partially supported by a grant from the NASA/NSF NStars 
initiative, administered by JPL, Pasadena, CA. KLC acknowledges support from a 
NSF Graduate Research Fellowship. This publication makes use of data products 
from the Two Micron All Sky Survey, which is a joint project of the University 
of Massachusetts and IPAC/CalTech, funded by NASA and the NSF; the NASA/IPAC 
Infrared Science Archive, which is operated by JPL/CalTech, under contract with 
NASA; and the SIMBAD database, operated at CDS, Strasbourg, France.

\begin{figure}
\plotone{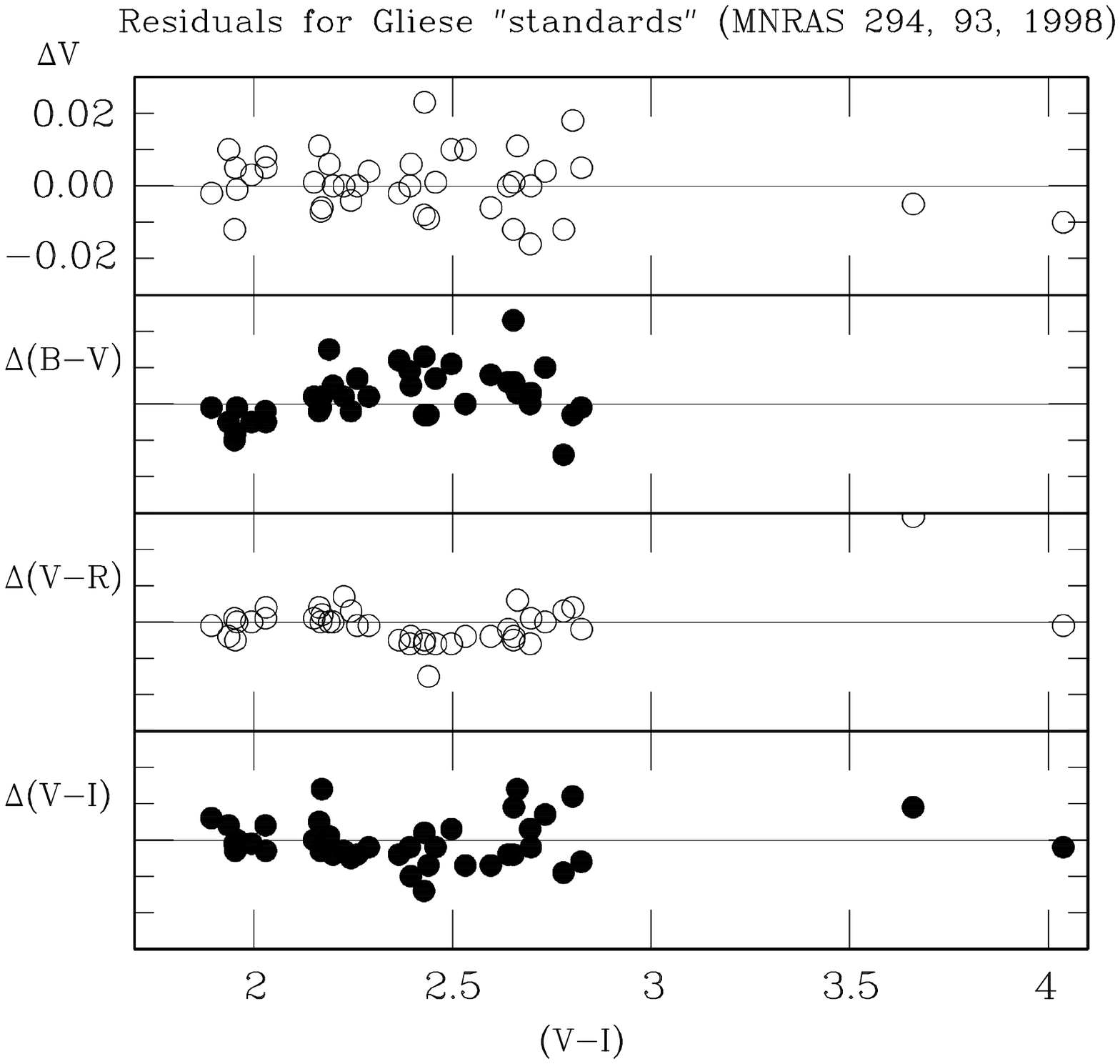}
\caption{ Photometric residuals for standard stars from the SAAO observations.}
\end{figure}

\begin{figure}
\plotone{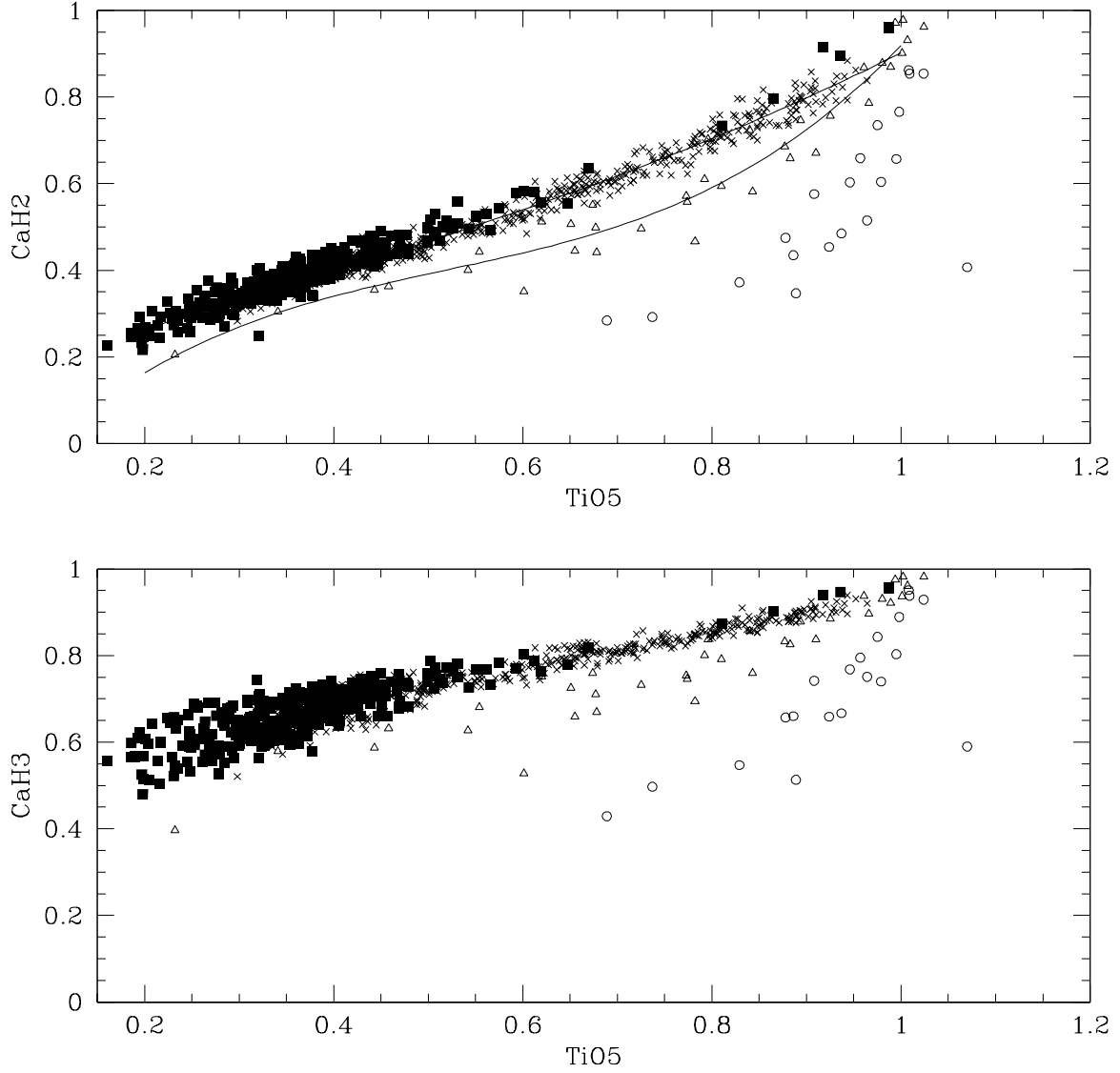}
\caption{ The TiO/CaH bandstrength measurements for the NLTT sample. The 
crosses plot reference data for disk dwarfs from the PMSU sample (PMSU1); the 
triangles and open circles plot data for intermediate and extreme subdwarfs 
(from Gizis, 1997). The solid lines in the upper diagram plot the mean 
relations for disk dwarfs and intermediate subdwarfs.
Filled squares mark measurements of the NLTT1 stars
listed in tables 2 and 3; apart from LP 381-49 (see text), 
all have bandstrengths consistent with near-solar 
abundance.}
\end{figure}

\begin{figure}
\plotone{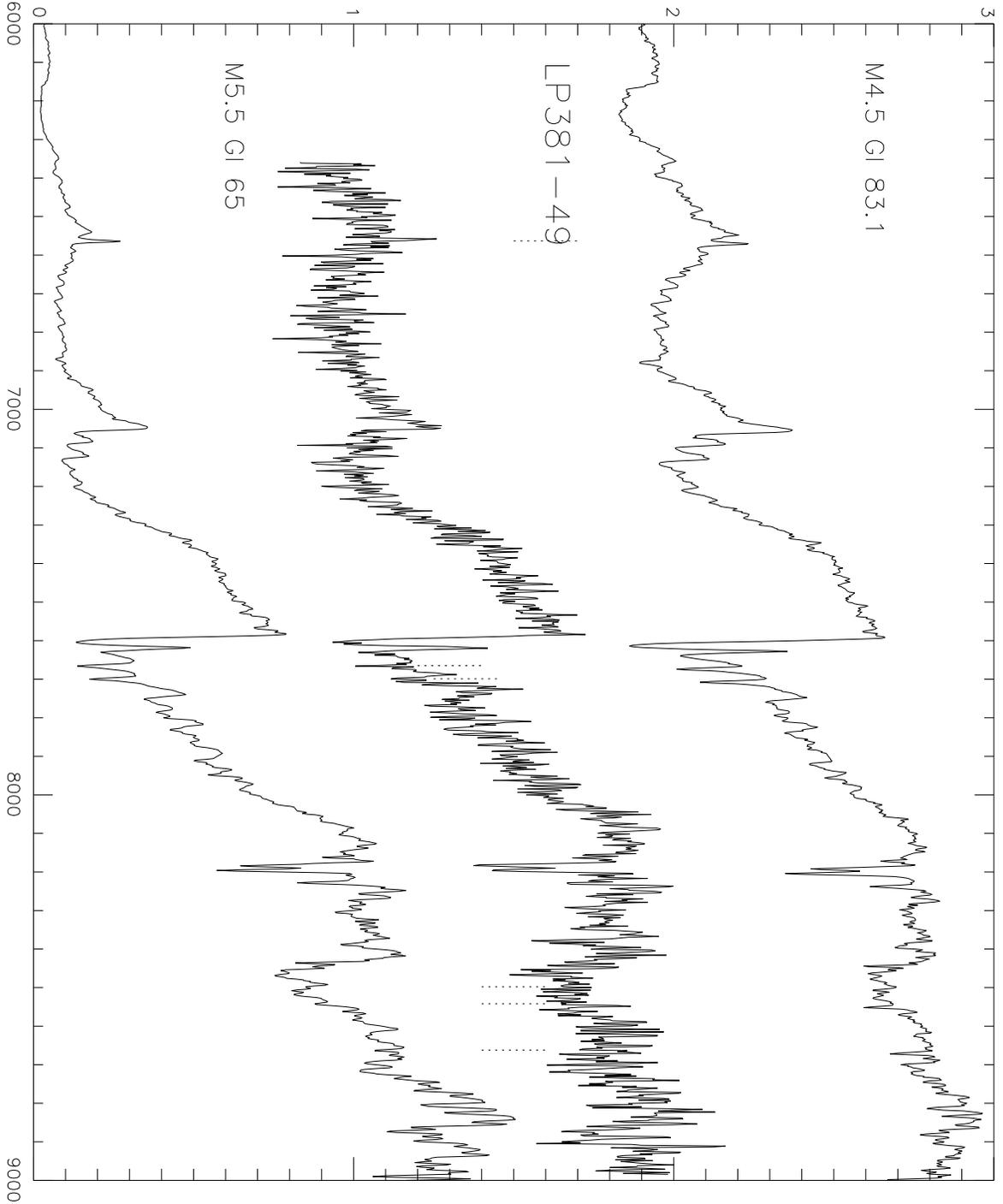}
\caption {Our observations of LP 381-49, a candidate intermediate subdwarf, compared 
to the spectral standards Gl 83.1 and Gl 65. The dotted vertical lines marks the 
location of H$\alpha$, KI and the Ca II triplet. }
\end{figure}

\begin{figure}
\plotone{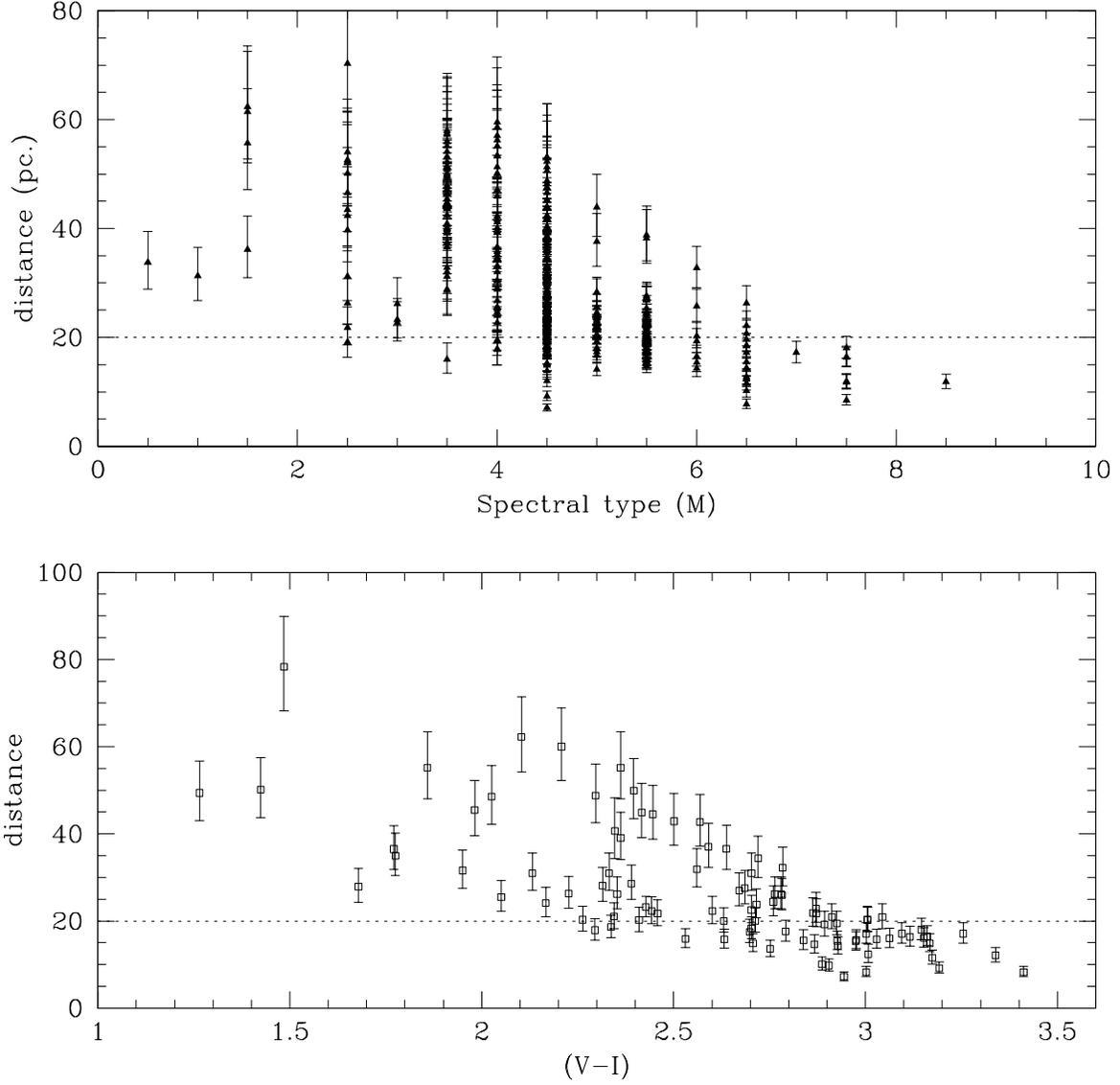}
\caption{ The distance distribution for the NLTT stars discussed in this paper.
The upper diagram shows the distances deduced for stars with spectroscopic 
observations; the lower plots data for the 96 stars with BVRI data. The dotted
line marks the formal distance limit for the present survey.}
\end{figure}

\begin{figure}
\plotone{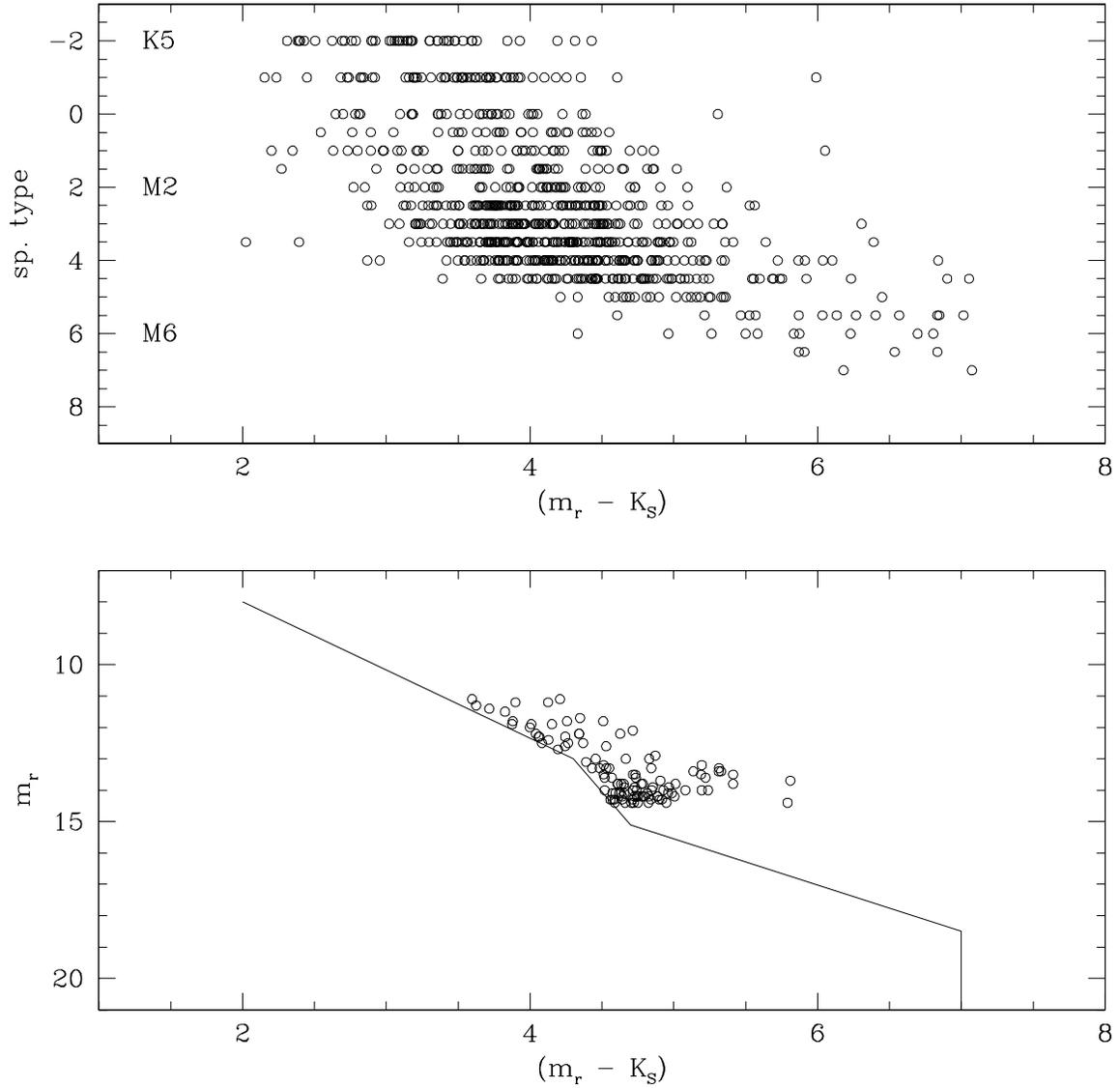}
\caption{ The upper panel shows the distribution of (m$_r$-K$_S$) colours as a 
function
of spectral type for PMSU stars within the area covered by the 2M2nd. The lower 
panel
plots the (m$_r$, (m$_r$-K$_S$)) distribution for the 112 stars in the NLTT 1 
sample
which still lack accurate follow-up photometry or spectroscopy. 
The solid lines outline our colour-magnitude selection criteria.}
\end{figure}

\begin{figure}
\plotone{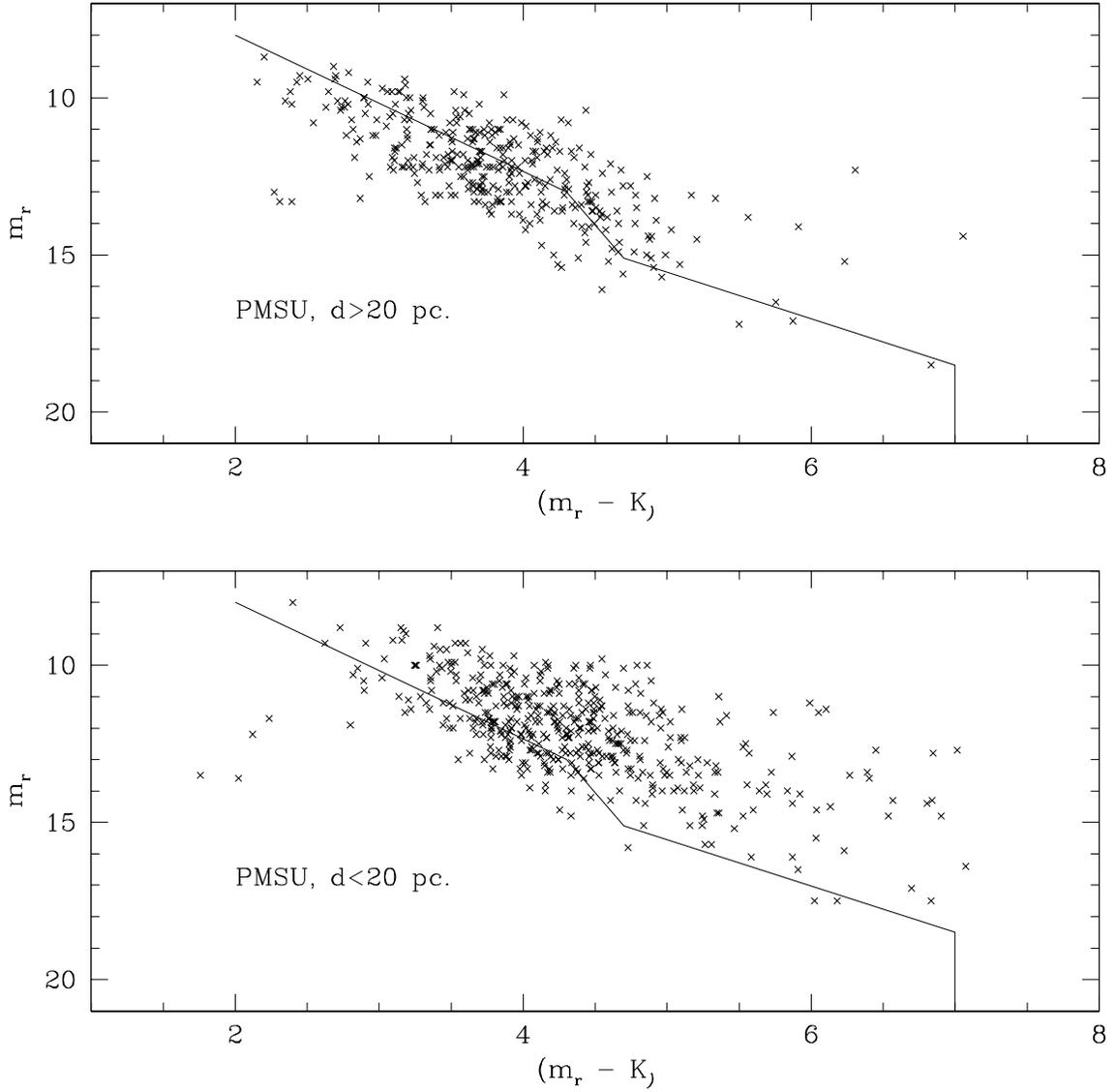}
\caption{ The (m$_r$, (m$_r$-K$_S$)) colour-magnitude diagrams for stars from 
the PMSU 2M2nd  dataset
which are also catalogued in the NLTT. The upper panel plots data for stars with 
distance
estimates, from PMSU, which exceed 20 parsecs; the lower panel plots data for 
stars within our 20-parsec limit.
In both cases, the solid lines mark the selection criteria adopted in paper I
to construct the NLTT1 sample.}
\end{figure}

\begin{figure}
\plotone{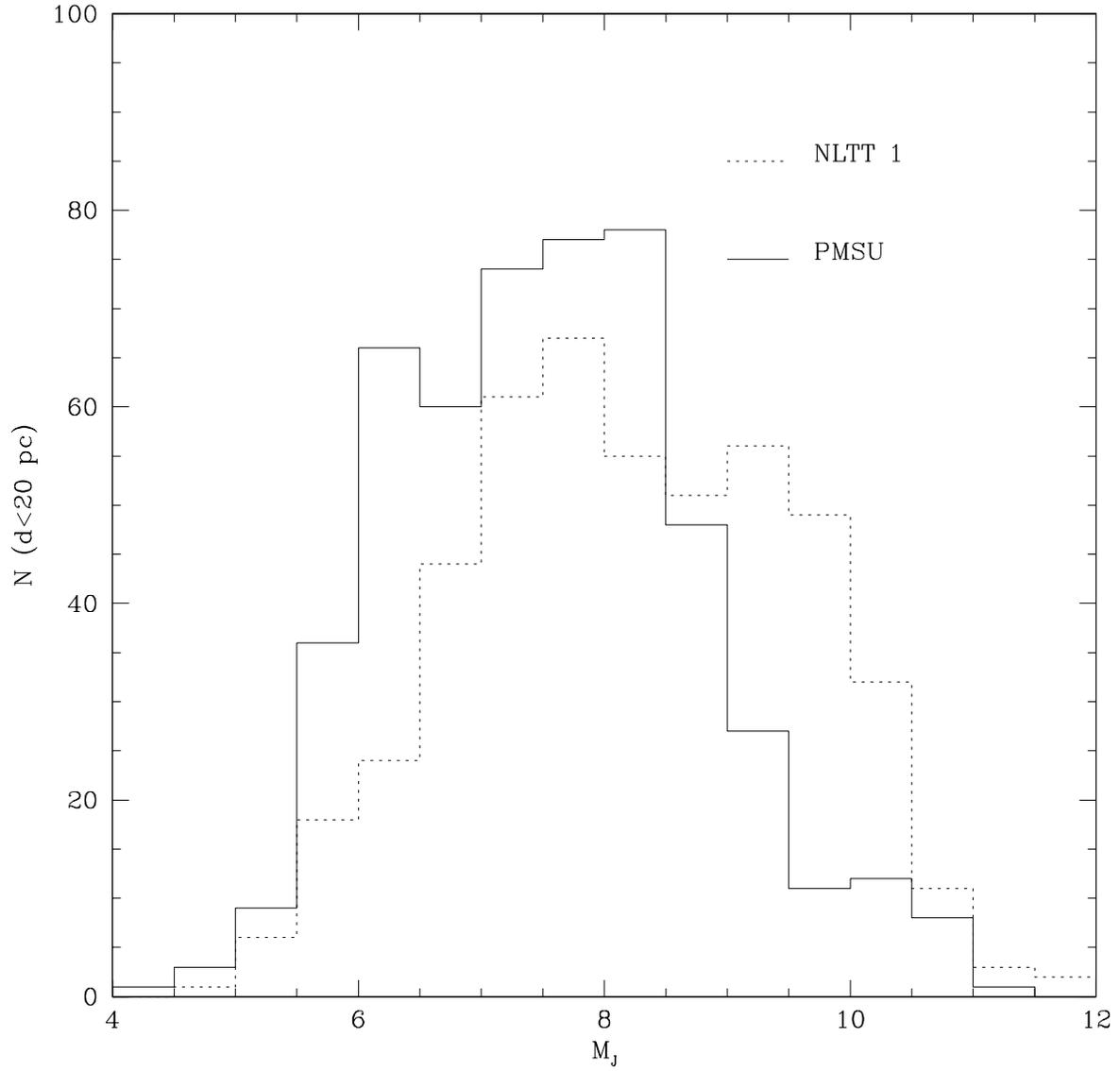}
\caption{ Additions to the nearby star census: the solid line plots the
the M$_J$ distribution for PMSU stars in the region covered by the 
2M2nd dataset with distances (from PMSU) $d < 20$ pc; 
the dotted line plots the same distribution for stars with
$d<20$-pc stars from the NLTT1 
sample. It is clear that the NLTT1 sample is contributing 
significant numbers of additional late-type M dwarfs to the current census.}
\end{figure}

\clearpage

\begin{deluxetable}{rrrrrrrrrc}
\rotate
\tabletypesize{\scriptsize}
\tablecolumns{10}
\tablenum{1}
\tablewidth{0pt}
\tablecaption{Ultracool NLTT1 dwarfs in Paper V}
\tablehead{\colhead{name} &
\colhead{ $\alpha$ (2000)} & \colhead{$\delta$} & \colhead{$m_r$} &
\colhead {J} &  \colhead {H} 
&\colhead {K$_S$} & \colhead{d} & \colhead{M$_J$} & \colhead {Notes }}
\startdata
LP 649-93 &   02:18:57.8 &  -06:17:49 & 19.2 &  12.920 &  12.186 &  11.860 & $36.6\pm5.0$ & 10.10 & M6, 1 \\
LP 944-20 &  03:39:35.2 & -35:25:44 & 17.5 &  10.748 & 10.017 &  9.525 & $5.0\pm0.1$ & 12.25 & M9.5, 2 \\ 
BD+34:1378B &  06:32:02.0 & 34:31:13&  14.7 &  10.694 & 10.132 &  9.876 & $29.4\pm1.4$ & 8.35 & 3 \\
LP 423-14 &   07:41:06.8 & 17:38:45 &  18.1 & 11.995 &  11.362 & 10.969 & $17.9\pm2.1$ & 10.70 & M7, 1 \\ 
LHS 2090 &  09:00:23.5 &  21:50:05 & 15.5 &   9.423 &  8.856 &  8.429 & $6.1\pm0.7$ & 10.50 & M6.5, 4 \\
LHS 2243 &  10:16:34.7 & 27:51:49 & 17.4 &  11.951 &  11.294&   10.946 & $14.1\pm1.7$ & 11.20 & M8, 5 \\ 
LP 213-67 &  10:47:12.6 & 40:26:43 & 16.3  & 11.417 & 10.777&  10.400 & $15.4\pm1.7$ & 10.50 & M6.5, 6 \\
LP 213-68 &  10:47:13.8 &  40:26:49 & 18.7  & 12.445 & 11.705 &  11.277 & $17.7\pm2.1$ & 11.20 & M8, 6 \\
\enddata
\tablecomments{ JHK$_S$ from 2MASS, $m_r$ from the NLTT \\
1. Sp. type from Paper V, distance from sp. type./M$_J$ relation. \\
2. Brown dwarf, distance based on trigonometric parallax (Tinney, 1996). \\
3. HD 257728B/G103-36B; Hipparcos parallax for primary. \\
4. Scholz \etal (2001), distance from sp. type./M$_J$ relation. \\
5. Kirkpatrick \etal (1995),  distance from sp. type./M$_J$ relation. \\
6. Gizis \etal (2000a),  distance from sp. type./M$_J$ relation. \\
7. Ruiz \& Takamiya (1995), distance from sp. type./M$_J$ relation. Henry \etal (2002)
estimate sp. type M4.5 and a distance of 20 pc. \\
}

\end{deluxetable}

\begin{deluxetable}{rrrrrrrrrc}
\tabletypesize{\scriptsize}
\tablecolumns{10}
\tablenum{2}
\tablewidth{0pt}
\tablecaption{Astrometry and photometry of spectroscopic targets}
\tablehead{\colhead{name} &
\colhead{ $\alpha$ (2000)} & \colhead{$\delta$} & \colhead{$m_r$} &
\colhead {J} &  \colhead {H} &\colhead {K$_S$} & \colhead{M$_J$} 
& \colhead{d (pc.)} & \colhead {Notes }}
\startdata
 LP 824-354   &  0: 5:29.8 & -25:31: 0. &    14.4 &  10.308 &   9.675 &   9.370 &   6.87 &   48.7$\pm$7.4 &    \\
 LP 824-355   &  0: 5:30.1 & -25:31: 6. &    13.2 &   9.435 &   8.773 &   8.459 &   6.65 &   36.0$\pm$5.5 &    \\
 LP 191-43$^{**}$    & 00:08:55.1 &  49:18:56  &    15.5 &  10.849 &  10.329 &   9.968 &   9.86 &   15.8$\pm$1.3 &    \\
  LP 644-53   & 00:09:59.8 &  -6:32:01  &    15.3 &  11.175 &  10.592 &  10.293 &   8.47 &   34.8$\pm$5.8 &    \\
  LP 880-607  & 00:11:58.9 & -29:32:50  &    15.2 &  11.084 &  10.540 &  10.211 &   7.53 &   51.3$\pm$7.4 & G267-49    \\
  LP 149-35   & 00:12:57.1 &  50:59:17  &    15.8 &  11.385 &  10.804 &  10.487 &   9.44 &   24.5$\pm$2.1 &    \\
  LP 880-754  & 00:15:02.3 & -27:02:44  &    14.5 &  10.688 &  10.069 &   9.828 &   6.75 &   61.4$\pm$9.4 & cpm G266-75   \\
  LP 824-286  & 00:17:32.6 & -24:31:06  &    15.1 &  11.237 &  10.583 &  10.318 &   8.23 &   39.9$\pm$6.7 &    \\
\enddata
\tablecomments{ \\
$^{**}$ - stellar systems with formal distance estimate $d < 20$ parsecs \\
$^*$ - stellar systems with formal distance estimates within $1\sigma$ of our 20 parsec limit \\
1. see Gizis \etal, 2000b, and Paper V \\
2. see also Paper III \\
3. also listed as APMPM J0150-3319, Scholz \etal, 2000. \\
4. LP 30-55 = G 245-40 is listed as GJ 3125 in the CNS3 and included in the PMSU survey. \\
5. LP 301-8 is listed as LP 301-63 in SIMBAD; photometry by Weis \& Hanson (1988) give
V=15.04, (V-R)$_C$=1.29, (V-I)$_C$=2.96. The star is a possible member of the
Hyades cluster.
6. LP 891-13 is a wide cpm companion of the white dwarf WD0443-275, see Silvestri \etal (2001). \\
7. DENIS-P J060254.2-091503, see Phan-Bao \etal, 2001. \\
8. LP 429-12/LHS 2215 - spectral type M6, Bessell, 1991. \\ 
9. LP 95-334 is a cpm companion of HD 110869, G5, $\pi_{Hip} = 22.10\pm0.81$ milliarcseconds. \\
10. see Ruiz \etal., 2001. \\
11. LP 617-34 is a wide cpm companion of the DA white dwarf, LP 617-35, Silvestri \etal, 2001. \\
12.  DENIS-P J140649.3-301828,  see Phan-Bao \etal, 2001. \\
13. DENIS-P J141206.9-041348,  see Phan-Bao \etal, 2001. \\
14. LHS 3002 is listed as GJ 3879B in the CNS3 and included in the PMSU survey. The cpm
companion is LHS 3001, for which we measure a photometric distance of 18.0$\pm$1.3 (Paper I)\\
15. 2MASSW J1504162-235556, see Reid \etal, 2002. \\
16. LP 759-25 = BRI B2202-1119, Kirkpatrick, Henry \& irwin, 1997. \\
17. see also Paper V
}
\end{deluxetable}

\begin{deluxetable}{rrrrrrrrrc}
\tabletypesize{\scriptsize}
\tablecolumns{10}
\tablenum{3}
\tablewidth{0pt}
\tablecaption{Narrowband indices for NLTT dwarfs }
\tablehead{\colhead{name} &
\colhead{ TiO5} & \colhead{CaOH} & \colhead{CaH1} &
\colhead {CaH2} &  \colhead {CaH3} 
&\colhead {TiO-a} & \colhead{Vo-a} & \colhead{H$\alpha$} & \colhead {Sp. Type }}
\startdata
 LP 824-354  &   0.44 &   0.48 &   0.89 &   0.48 &   0.73 &   1.46 &   1.92 &   0.94 & M3.5 \\
 LP 824-355  &   0.50 &   0.53 &   0.86 &   0.58 &   0.81 &   1.37 &   1.89 &   0.94 & M3.0 \\
 LP 191-43   &   0.27 &   0.25 &   0.71 &   0.29 &   0.56 &   2.25 &   2.02 &   1.22 &   M5.5 \\
  LP 644-53  &   0.37 &   0.39 &   0.80 &   0.39 &   0.65 &   1.69 &   2.00 &   1.58 &   M4.5 \\
  LP 880-607 &   0.44 &   0.40 &   0.80 &   0.45 &   0.71 &   1.47 &   1.94 &   0.95 &   M3.5 \\
  LP 149-35  &   0.32 &   0.38 &   0.83 &   0.38 &   0.67 &   1.95 &   2.01 &   1.84 &   M5.0 \\
  LP 880-754 &   0.65 &   0.57 &   0.84 &   0.55 &   0.78 &   1.19 &   1.96 &   0.96 &   M1.5 \\
  LP 824-286 &   0.42 &   0.41 &   0.83 &   0.41 &   0.71 &   1.52 &   1.96 &   0.99 &   M3.5 \\
\enddata
\tablecomments{ 
}
\end{deluxetable}

\begin{deluxetable}{rrrrrrrrrrrrrrrrc}
\rotate
\tabletypesize{\scriptsize}
\tablecolumns{17}
\tablenum{4}
\tablewidth{0pt}
\tablecaption{Photometry of NLTT dwarfs}
\tablehead{\colhead{name} &
\colhead{ $\alpha$ (2000)} & \colhead{$\delta$} & \colhead{V} &
\colhead { (B-V)} & \colhead { (V-R)} & \colhead { (V-R)} & \colhead { n$_{obs}$} & 
\colhead {$\sigma_V$} &\colhead {$\sigma_{B-V}$} &\colhead {$\sigma_{V-R}$} &\colhead {$\sigma_{V-I}$} &
\colhead {J} &  \colhead {H} 
&\colhead {K$_S$} & \colhead{M$_J$} & \colhead{d (pc.)}}
\startdata
 644- 39     & 0:06:13.0& -2:32:10&  14.622& \nodata&   1.203&   2.782&  2& 14&  0&  4& 10&  10.236&   9.613&   9.312&   8.17$\pm$0.25&   25.9 \\
 880-769     & 0:16:03.3&-30:58:42&  13.719& \nodata&   0.988&   2.104&  1&\nodata &\nodata &\nodata &\nodata &  10.449&   9.890&   9.605&   6.48$\pm$0.25&   62.2 \\
 584- 94*    & 0:17:40.6& -1:22:40&  12.886&   1.457&   1.049&   2.390&  1&\nodata &\nodata &\nodata &\nodata &   9.261&   8.567&   8.370&   6.98$\pm$0.25&   28.6 \\
 825- 48     & 0:20:50.5&-25:34:05&  12.551& \nodata&   0.953&   1.982&  1&\nodata &\nodata &\nodata &\nodata &   9.518&   8.821&   8.614&   6.23$\pm$0.25&   45.5 \\
 705- 15     & 0:21:39.4& -9:00:24&  13.939& \nodata&   1.175&   2.703&  2& 14&  0&  6&  7&   9.732&   9.159&   8.879&   7.97$\pm$0.25&   22.5 \\
 705- 22     & 0:24:25.1&-12:17:24&  12.245&   1.497&   1.011&   2.227&  1&\nodata &\nodata &\nodata &\nodata &   8.808&   8.191&   7.949&   6.71$\pm$0.25&   26.3 \\
 645- 48     & 0:34:38.9& -2:25:00&  11.952&   1.496&   0.999&   2.167&  1&\nodata &\nodata &\nodata &\nodata &   8.569&   7.948&   7.685&   6.66$\pm$0.25&   24.1 \\
 705- 65     & 0:35:38.0&-10:04:18&  11.831&   1.523&   1.025&   2.263&  1&\nodata &\nodata &\nodata &\nodata &   8.314&   7.737&   7.519&   6.77$\pm$0.25&   20.3 \\
\enddata
\tablecomments{ 
Column lists the star name from the NLTT; columns 2 and 3 give the position from 2MASS;\\
columns 4-7 list the photometry, column 8 the number of observations, and, for multiple
observations,  columns 9-12 give the formals residuals about the mean, in millimags; \\
columns 13-15 list the 2MASS photometry; column 16 gives the derived absolute magnitude
and uncertainty; and column 17 gives the corresponding distance in parsecs. \\
Notes to individual stars: \\
1. G 274-24: both 2MASS and optical photometry include close binary companion, sep. $\sim2\farcs$. 
Jao \etal (2003) list this source as GJ 2022AC, and give $\delta V = 0.06$ mag. Magnitudes
adjusted by +0.75 mag. LP 939-44 (GJ 2022B) is 2MASS 01243060-3355014, J=10.55, H=10.009, 
K=9.682.
2. LP 793-24: optical photometry includes LP 793-25, sep. 4.5\farcs, $\delta m \sim 0.2$ mag; 
V magnitude adjusted by +0.75 mag. \\
3. LP 798-78: optical photometry includes 2 background stars, sep. $\sim 5\farcs$, 
$\delta m \sim 1$ and 2 mag. V magnitude adjusted by +0.35 mag. \\
4. LP 693-21: optical photometry includes a background star, sep. 7\farcs, $\delta m \sim 1.5$ mag; 
V magnitude adjusted by +0.25 mag. \\
}

\end{deluxetable}

\end{document}